\documentclass[11pt,superscriptaddress]{article}

\usepackage[a4paper, total={5.8in, 9in}]{geometry}
\usepackage[indent,skip=4pt]{parskip}
\usepackage{amsmath,amssymb,amsfonts,amsthm,mathrsfs,mathtools,cases}
\usepackage{dsfont}
\usepackage{authblk}
\usepackage[colorlinks=true,linkcolor=blue, citecolor=blue, bookmarks]{hyperref}
\usepackage{cleveref}
\usepackage{microtype}
\usepackage{comment}
\usepackage{graphicx,adjustbox} 
\usepackage{subfig}
\usepackage{wrapfig}
\usepackage[font=small,labelfont=bf]{caption}
\usepackage[usenames,dvipsnames]{xcolor}  
\usepackage{booktabs}
\usepackage{braket}
\usepackage[nottoc]{tocbibind}
\usepackage{nicematrix}
\usepackage{tikz}
\usepackage{pgfplots}
\usepackage{cite}
\pgfplotsset{compat=1.18}
\usepackage{enumitem}

\usetikzlibrary{matrix}
\usetikzlibrary{decorations.markings}
\usetikzlibrary{patterns}
\usetikzlibrary{arrows.meta}

\numberwithin{equation}{section}

\newcommand{\ZZ}{\mathbb Z}

\newcommand{\boldg}{\mathbf g}

\newcommand{\Zng}{\mathcal Z_n(\boldg)}

\DeclareMathOperator{\tr}{tr}

\newcommand{\lr}[1]{\left( {#1} \right)}
\newcommand{\slr}[1]{\left[{#1} \right]}

\crefname{figure}{Fig.}{Figs.}
\crefname{equation}{Eq.}{Eqs.}
\crefname{section}{Sec.}{Secs.}
\crefname{appendix}{Appendix}{Appendices}
  
\colorlet{darkerblue}{MidnightBlue!70!black}
\colorlet{lightblue}{blue!70!white}

\hypersetup{
    colorlinks,
    linkcolor={darkerblue},
    citecolor={darkerblue},
    urlcolor={darkerblue},
    backref=true
}

\usepackage[framemethod=TikZ]{mdframed}

\title{\bf Entanglement asymmetry in CFT  with boundary symmetry breaking}

\author{Michele Fossati$^1$, Colin Rylands$^1$ and Pasquale Calabrese$^{1,2}$}

\date{}

\begin{document}
\maketitle
{\small
\vspace{-5mm}  \ \\
{$^{1}$}  SISSA and INFN Sezione di Trieste, via Bonomea 265, 34136 Trieste, Italy\\[-0.1cm]
\medskip
{$^{2}$}  International Centre for Theoretical Physics (ICTP), Strada Costiera 11, 34151 Trieste, Italy\\[-0.1cm]
\medskip
}

\begin{abstract}
We examine the behavior of the entanglement asymmetry in the ground state of a (1+1)-dimensional conformal field theory with a boundary condition that explicitly breaks a bulk symmetry. Our focus is on the asymmetry of a subsystem $A$  originating from the symmetry-breaking boundary and extending into a semi-infinite bulk.
By employing the twist field formalism, we derive a universal expression for the asymmetry, showing that the asymptotic behavior for large subsystems is approached algebraically, with an exponent which is twice the conformal dimension of a boundary condition-changing operator.
As a secondary result, we also establish a similar asymptotic behavior for the string order parameter. Our exact analytical findings are validated through numerical simulations in the critical Ising and 3-state Potts models.
\end{abstract}

\tableofcontents
\newpage

\section{Introduction}
The most widely used approach to symmetry and symmetry breaking in many body quantum systems is through an appropriate local order parameter~\cite{Sachdev:2011fcc}.  A symmetry is said to be broken in a particular state if the expectation value of the order parameter is non-zero. This approach, despite its many successes, contains some important and consequential drawbacks.  Aside from the difficulty of appropriately identifying and then actually calculating the value of the order parameter,  the method provides no determination on whether a symmetry is broken or not if the order parameter is zero.  Thus,  an order parameter cannot always distinguish between symmetry breaking and symmetry preserving states.  The problem is exacerbated if one is interested in the symmetry properties of an extended subsystem, in which case the spatially averaged order parameter may vanish despite being locally non zero.  

To surmount these issues and study symmetry breaking at the subsystem level, one can instead make use of the \emph{entanglement asymmetry}~\cite{Ares:2022koq}.  This observable independent quantity,  allows one to study symmetry breaking directly at the level of the reduced density matrix of the subsystem without recourse to any order parameter and faithfully distinguishes between symmetry breaking and symmetry preserving states.  It has been studied in a wide range of systems at equilibrium; for ground states with explicit~\cite{Murciano:2023qrv, Ferro:2023sbn, Fossati:2024xtn,Lastres:2024ohf }
and spontaneous symmetry breaking~\cite{CapizziMazzoniIsing:2023}, in matrix product states~\cite{CapizziVitaleMPS:2023}, excited states of conformal field theories~\cite{Chen:2023gql,Benini:2024xjv}, mixed states in free models~\cite{Ares:2024nkh} and Haar random states~\cite{Ares:2023ggj}. It is also applicable out of equilibrium and has been used to reveal several exotic phenomena such as the quantum Mpemba effect~\cite{Ares:2022koq,Ares:2023kcz,Bertini:2023ysg,Rylands:2023yzx,Khor:2023xar,Murciano:2023qrv,Yamashika:2024hpr,Turkeshi:2024juo,Liu:2024kzv,Liu:2024uqf,Chalas:2024wjz,Klobas:2024mlb,Caceffo:2024jbc, Foligno:2024jpq,Rylands:2024fio,Yamashika:2024mut,Maric:2024hzo}, which has been observed experimentally~\cite{Joshi:2024sup}. The same quantity was introduced independently in resource theory, under the name \emph{Holevo asymmetry}~\cite{Gour:2009abc,Marvian:2014awa}, and algebraic quantum field theory, as the \emph{entropic order parameter}~\cite{Casini:2019kex,Casini:2020rgj,Magan:2021myk}. 
Such studies have focused, for the most part, on the global breaking of symmetry. A thus far  under investigated scenario is the case wherein the symmetry is instead broken \textit{locally}, either by a defect or the choice of boundary conditions.  The effects of local symmetry breaking can, however, be just as interesting and consequential, particularly in the case of one-dimensional critical systems, which we focus on here. 

\begin{figure}
    \centering
    \begin{tikzpicture}[scale=0.8]
        \draw (0,0) -- (5,0);
        \draw[dashed] (5,0) -- (6,0);
        \draw[orange, very thick]  (0,-0.2) --  (0,0.2) ;
        \node[orange, left] at (0,0) {$a$};
        \fill (2,0) circle[radius=1.5pt] node[above] {$\ell$}; 
        \node[above] at (0,0.1) {$0$}; 
        \node[above] at (6,0) {$+\infty$}; 

        \draw[decorate,decoration={brace,mirror,amplitude=5pt}] (0,-0.4) -- (2,-0.4) node[midway,below=4pt] {$A$};
    \end{tikzpicture}
    \caption{We consider a one dimensional, $G$ symmetric CFT on the half line subject to a conformal boundary condition, $a$ on the left edge which breaks the $G$ symmetry. We invetigate properties of the subsystem $A$ which extend from the boundary a distance $\ell$  into the bulk.  }\label{fig:schematic}
\end{figure}
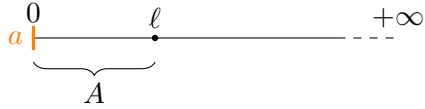

In this paper,  we employ the entanglement asymmetry  to study the ground state properties of (1+1)-dimensional conformal field theories in the presence of symmetry breaking boundary conditions.  We consider the ground state of a conformal field theory, $T$ on the half-infinite line $x \in [0, \infty)$, with a boundary condition, $a$, at $x=0$. Unless otherwise stated our subsystem of interest, $A$ shall extend from the boundary such that $A=[0,\ell]$, see Fig.~\ref{fig:schematic}. The CFT is assumed to possess a global symmetry, $G$, which is however broken by the boundary condition $a$.
We use the methods of boundary conformal field theory to derive analytic expressions for the entanglement asymmetry, as well as a related quantity known as the full counting statistics (aka string order parameter),   and unveil an interesting scaling relationship between the symmetry breaking in the subsystem and the dimension of the boundary condition changing operators of the theory.  We then benchmark these against numerical results obtained using tensor network methods on lattice systems, specifically the Ising spin chain and the 3-state Potts model. 

\subsection{Entanglement Asymmetry}

The entanglement asymmetry is an information-theory-based quantity that measures how much a state breaks a certain symmetry group $G$ \cite{Marvian:2014awa, Ares:2022koq,Casini:2019kex}. Assuming that the group is represented by operators $U_g, g\in G$ that factorize over the subsystems' Hilbert spaces as $U_g = U_{A,g} \otimes U_{\bar A, g}$,  the method entails comparing the reduced density matrix $\rho_A$ of a certain state $\rho$ with its symmetrized version $ \rho_{A,G} $.  This is defined as~\cite{CapizziVitaleMPS:2023},
\begin{equation}\label{eq:def-rhoG}
    \rho_{A,G} = \frac{1}{|G|} \sum_{g\in G} U_{A,g} \rho_A U_{A,g}^\dagger.
\end{equation}
Here,  we have restricted to $G$ being a finite group,  but continuous Lie groups can also be treated, see~\cite{Fossati:2024xtn} for the definition of $\rho_{A,G}$ in that case, as can space-time symmetries~\cite{Klobas:2024png}.  The entanglement asymmetry is  then defined as the quantum relative entropy between $\rho_A$ and $\rho_{A,G}$~\cite{Ares:2022koq},
\begin{equation}
    \Delta S_A = S(\rho_A || \rho_{A,G}) = \tr[ \rho_A ( \log \rho_A - \log \rho_{A,G} ) ].
\end{equation}
The properties of relative entropies mean that $ \Delta S_A\geq 0$ with equality if and only if $\rho_A=\rho_{A,G}$, i.e.  if the state is symmetric.   
Moreover, it can be shown that $\rho_{A,G}$ is the symmetric state which minimizes the relative entropy with  $\rho$~\cite{Gour:2009abc},  thereby justifying the choice~\eqref{eq:def-rhoG} as opposed to some other symmetric state.   Thus, the entanglement asymmetry quantifies how far a state is from being symmetric by measuring how different it is from its symmetrized version. The larger the value of $\Delta S_A$ the more the state breaks the symmetry.    Note that, by construction, any symmetric state obeys $\rho_{A,G}=\rho_A$ and hence we automatically overcome the pitfall of the order parameter.  The tradeoff of this approach is that, in general,  $\Delta S_A$ is a much more complicated quantity to compute, however,  there are now numerous methods to achieve this.

Using the cyclicity of the trace and the properties of $\rho_{A,G}$ one can show that $\Delta S_A$ can expressed as the difference of the von Neumann entropies of the two states, 
\begin{equation}
\Delta S_A=S(\rho_{A,G})-S(\rho_A),
\end{equation}
where $S(\rho)=-\tr [\rho\log (\rho)]$.
As a result, the entanglement asymmetry can be obtained as the limit $n \to 1$ of the R\'enyi entanglement asymmetry,
\begin{equation}
    \Delta S^{(n)}_A = \frac{1}{1-n} \log \frac{\tr \rho_{A,G}^n}{\tr \rho_A^n}.
\end{equation}
Upon  substituting \eqref{eq:def-rhoG} in the above one gets
\begin{equation}\label{eq:asymmetry-sum-group}
    \Delta S^{(n)}_A = \frac{1}{1-n} \log \slr{ \frac{1}{|G|^{n-1}} \sum_{\substack{g_1, \dots, g_n \in G \\ \prod_j g_j = e}} \mathcal Z_n(g_1, \dots, g_n) } ,
\end{equation}
where $e$ is the identity element of the group and
\begin{equation} \label{eq:def-normalized-charged-moment}
    \mathcal Z_n(g_1, \dots, g_n) = \mathcal Z_n(\boldg) = \frac{\tr(\rho_A U_{A, g_1} \dots \rho_A U_{A,g_n}) }{\tr \rho_A^n},
\end{equation}
which we refer to as the  \emph{normalized charged moment}. Note that for the purposes of evaluating the entanglement asymmetry, it is the neutral charged moment, where $\prod_j g_j=e$, which is used. This property will be crucial in determining the correct behaviour of $\Delta S_A$. 

\subsection{Full counting statistics and the string order parameter}
The normalized charged moments are quantities of interest in their own right.  
In particular, the simpler case  $n=1$  has been widely discussed in the literature. 
For Lie groups, this case is also known as the full counting statistics and carries special importance.
It is the moment generating function of the generators of the group restricted to the subsystem. 
For discrete groups,  this interpretation does not persist, however, it can nevertheless unveil information about the symmetry breaking properties of the state.  
This is the spirit of an alternative approach to boundary symmetry breaking~\cite{Bonsignori:2023tpk}, which is motivated by the concept of string order parameter~\cite{Kennedy:1992ifl,Kennedy:1992abc}.   
According to Ref.~\cite{Bonsignori:2023tpk}, one takes $A$ to be the full system and establishes that the state is symmetric if and only if
\begin{equation}
\log{ |\mathcal Z_1(g)|}=0,~~\forall g\in G, 
\end{equation}
whereas the symmetry is broken if it attains a non zero value. 
Similarly to $\Delta S_A$, this measure faithfully detects symmetry breaking of the full state and works for both continuous and discrete groups. 
Notice that, despite its name, it is a non-local quantity and indeed is often referred to as a non-local order parameter.
It is certainly simpler to calculate than entanglement asymmetry. 
However, although it has been studied in both equilibrium~\cite{Bonsignori:2023tpk} and out-of-equilibrium settings~\cite{Barad:2024hea}, it remains less powerful as a diagnostic tool than the entanglement asymmetry. 
In particular, its effectiveness as an indicator of symmetry breaking for mixed states and, consequently, for subsystems is less conclusive. 
Nevertheless, we will investigate the string order parameter, $\log{ |\mathcal Z_1(g)|}$, restricted to the subsystem $A$ and compare the obtained results with those from the asymmetry.

\subsection{Main results}
Let us briefly present our main results with their derivation following in the next section. 
We consider a boundary condition $a$ at $x=0$ and the subsystem $A=(0,\ell]$. 
We find that the string order parameter admits the scaling form
\begin{equation}
\mathcal{Z}_1(g)\propto \ell^{-2 h_\mu- h_{\Psi_{(ga,a)}}}
    \label{stringintro}
\end{equation}
where $h_\mu$ is the scaling dimension of the (diagonal) disorder operator associated to the group element $g \in G$ and $h_{\Psi_{(ga,a)}}$ the conformal dimension of the boundary-condition-changing operator $\Psi_{(ga,a)}$ that changes the boundary condition from $a$ to $ga$.

For a finite group $G$ of cardinality $|G|$, we find that the entanglement asymmetry scales as
\begin{equation}
    \Delta S^{(n)}_A = \log |G| - c_n \ell^{-2h^\mathrm{min} } + o( \ell^{-2h^\mathrm{min} })
     \label{asyintro}  
\end{equation}
with $h^\mathrm{min}$ being the minimal dimension of the boundary-condition-changing operators within the group $G$, i.e., 
\begin{equation}
    h^\mathrm{min} = \min_{g \in G,g\neq e} h_{\Psi_{(ga,a)}}. 
\end{equation}
The constant $c_n$ is just a non-universal amplitude.  

Note that, as the string order parameter approaches zero for large subsystems, cf. Eq. \eqref{stringintro}, one might be tempted to infer that the subsystem becomes increasingly symmetric as its size grows. However, the result for the entanglement asymmetry in Eq. \eqref{asyintro} demonstrates otherwise, highlighting the greater effectiveness of entanglement asymmetry in detecting symmetry breaking at the subsystem level. Moreover, the string order parameter on the subsystem remains nonzero even in the presence of a symmetry preserving boundary, $h_{\Psi_{(ga,a)}}=0$.

We emphasize also that, while  the traditional order parameter approach may yield a nonzero value in certain cases, its value decays algebraically away from the symmetry breaking boundary. Moreover, its value provides no direct information on the symmetry being broken. In contrast, the asymmetry does not decay as the subsystem increases correctly reflecting the fact that the symmetry is not restored and its value indeed provides  information on the broken symmetry group.

\subsection{Organization of the manuscript}
The manuscript is organized as follows. 
In Sec. \ref{main} our main results anticipated above for the string order parameter and the asymmetry are derived. 
Sec. \ref{numerics} is devoted to provide concrete examples of the general results and to their numerical verification in both the critical Ising and 3-state Potts models.
Sec. \ref{concl} reports our conclusions and outlooks. 

\section{Derivation of the main results}
\label{main}

In the two following subsections we derive the anticipated main results for the string order parameter and entanglement asymmetry.

\subsection{String order parameter}

We begin by considering the string order parameter in the ground state of our chosen CFT, $T$.  This can be expressed as
\begin{equation}
\mathcal{Z}_1(g)=\lim_{\beta\to\infty}\frac{\tr[U_{A,g}e^{-\beta H_a}]}{\tr[e^{-\beta H_a}]}
\end{equation}
where $H_a$ is the Hamiltonian of the theory on the half line subject to boundary condition $a$ at $x=0$.  As is standard,  our theory,  $T$,  then lives on the right half complex plane, coordinated by $z$,  with a certain line operator, corresponding to $g$, inserted along ${\rm{Re}}(z)=(0,\ell]$. In addition, the points $z\to \pm i\infty$  should be identified.  
Recall that, in order to have a well defined insertion of a line operator, some conditions at its endpoints need to be specified. These consist of declaring on which defect operator (also called the \emph{disorder operator} or \emph{twist field}) the line ends in the bulk, and with which junction the line is connected to the boundary. We call the defect operator $\mu$ and assume that it is a diagonal primary operator, that is $h_\mu = \bar h_\mu$.

It is useful at this point, to introduce a diagrammatic representation of our quantities of interest.  In particular,  we express  the expectation value of $g$ in the groundstate in the following way
\begin{equation}
   \mathcal{Z}_1(g)= 
    \left \langle 
    \tikz[baseline=.1ex,scale =0.7]{
    \draw[dashed] (0,-1) rectangle (1.7,1);
    \draw[very thick, orange] (0,-1) -- node[midway, left] {$a$} (0,1) ;
    \draw[thick, blue] (0,0) -- node[midway, above] {$g$} (1,0);
    \fill (1, 0) circle(2pt) node[right] {$\mu$};
    \fill (0, 0) circle(2pt) ;
    }\,
    \right \rangle,
\end{equation}
where on the right hand side, the orange line indicates the boundary condition $a$, at ${\rm{Re}}(z)=0$, the dashed lines imply that the theory lives on the full right half plane,  the blue line denotes the line operator and the black dots denote local defect operators or junctions which are inserted where the line operators terminates. Throughout this work we are considering the case where the bulk theory is $G$-symmetric, meaning that the line operator is topological.  In the diagrammatic notation, this means that the blue line can be completely deformed to any shape provided its endpoints remain fixed and no other operators or boundaries are encountered.

We first consider the case in which the boundary condition is $G$-invariant, and we denote it with $i$.  In that case,  the junction between the boundary and the line operator is also topological\cite{Shao:2023gho}.  This is because the line can be deformed in the bulk and made to act on the boundary.  If the boundary is $G$-symmetric, then the line can disappear on the boundary. Therefore, the position of the junction can be moved without changing the  expectation value.   This is represented diagrammatically as, 
\begin{equation}
    \tikz[baseline=.1ex,scale =0.7]{
    \draw[very thick, orange] (0,-1) -- node[midway, left] {$i$} (0,1) ;
    \draw[thick, blue] (0,-0.5) -- node[midway, above] {$g$} (2,-0.5);
    \fill (2, -0.5) circle(1.5pt) node[right] {$\mu$};
    }
    =
     \tikz[baseline=.1ex,scale =0.7]{
     \draw[very thick, orange] (0,-1) -- node[midway, left] {$i$} (0,1) ;
    \draw[thick, blue] plot[smooth, tension=1] coordinates {(0,-0.5) (0.4,-0.5) (0.4,0.5) (2,-0.5)};
    \fill (2, -0.5) circle(1.5pt) node[right] {$\mu$};
    }
    = 
    \tikz[baseline=.1ex,scale =0.7]{
    \draw[very thick, orange] (0,-1) -- node[midway, left] {$i$} (0,1) ;
    \draw[thick, blue] (0,0.5) -- node[midway, above right] {$g$} (2,-0.5) ;
    \fill (2, -0.5) circle(1.5pt) node[right] {$\mu$};
    } ~.
\end{equation}
Accordingly,  the position of the junction cannot enter in any expectation value.  Note however, that the position of the defect operator in the bulk, is not topological and so cannot be moved without changing $\mathcal{Z}_1(g)$.   We can then use the doubling trick \cite{Cardy:1984bb} to evaluate the expectation value.  In particular, we map $T$ to a chiral theory $T_{\rm c}$ which lives on the full complex plane. Under this mapping anti holomorphic fields on the right half plane get mapped to the left half plane $\bar{\mu}(z)\to \bar{\mu}_{\rm c}(-\bar{z})$ while the holomorphic parts remain in the right half plane  $\mu(z)\to \mu_{\rm c}(z)$. We assume that $\mu_\mathrm{c}$ and $\bar{\mu}_{\mathrm c}$ have the same conformal dimension, denoted by $h_\mu$. This then maps the expectation value in the boundary CFT into a chiral expectation value on the complex plane. Following this we obtain 
\begin{equation} \label{eq:expectation-value-line-invariant-boundary}
    \mathcal{Z}_1(g)  =  \left \langle \,
    \tikz[baseline=.1ex,scale =0.7]{
    \draw[dashed] (-1.5,-1) rectangle (1.5,1);
    \draw[dotted] (0,-1) -- (0,1);
    \fill (1, 0) circle(1.5pt) node[below] {$\mu_\mathrm c$};
    \fill (-1, 0) circle(1.5pt) node[below] {$\bar{ \mu}_\mathrm{c} $};
    }
    \, \right \rangle_\mathrm{c}
    \propto \ell^{-2 h_\mu}.
    \end{equation}
where $\langle\cdot\rangle_{\rm c}$ refers to the expectation value in the chiral theory $T_{\rm c}$. Here we see that the scaling behaviour depends explicitly on the dimension of the defect operator which terminates the line operator. If we were to consider instead a finite  system such that the line operator ended on another $G$-symmetric boundary then we would obtain that $\mathcal{Z}_1(g)=1$,  in line with the fact that the ground state would be symmetric. 

If the boundary condition is not $G$-invariant, then the junction cannot be topological. This is because moving the line in the bulk and making it act on the boundary will change the boundary, and thus produce a boundary changing operator which we denote by $\Psi_{(ga,a)}$. This is depicted in terms of diagrams as
\begin{align} \label{eq:move-line-change-boundary}
    \tikz[baseline=.1ex,scale =0.7]{
    \draw[very thick, orange] (0,-1) -- node[midway, left] {$a$} (0,1) ;
    \draw[thick, blue] (0,-0.5) -- node[midway, above] {$g$} (2,-0.5);
    \fill (2, -0.5) circle(1.5pt) node[right] {$\mu$};
    \fill (0,-0.5) circle (2pt);
    }
    =
     \tikz[baseline=.1ex,scale =0.7]{
     \draw[very thick, orange] (0,-1) -- node[midway, left] {$a$} (0,1) ;
    \draw[thick, blue] plot[smooth, tension=1] coordinates {(0,-0.5) (0.2,-0.3) (0.45,0.5) (2,-0.5)};
    \fill (2, -0.5) circle(1.5pt) node[right] {$\mu$};
    \fill (0,-0.5) circle (2pt);
    }
    =
    \tikz[baseline=.1ex,scale =0.7]{
    \draw[very thick, orange] (0,-1) -- node[midway, left] {$a$} (0,-0.5) -- node[midway, left] {$ga$} (0,0.5) -- node[midway, left] {$a$} (0,1) ;
    \draw[thick, blue] (0,0.5) -- node[midway, above right] {$g$} (2,-0.5) ;
    \fill (0,-0.5) circle (2pt) node[right] {$\Psi_{(ga, a)}$};
    \fill (2, -0.5) circle(2pt) node[right] {$\mu$};
    }.
\end{align}
In this sense the junction ``hides a boundary changing operator'' with a positive conformal dimension. This was pointed out in the case of the Ising CFT in \cite{Watts:2000kp}. Notice that the topmost junction appearing in the RHS of \cref{eq:move-line-change-boundary} is instead topological, as it can be further moved away. The position of this latter junction, as in the case discussed above, will then not enter in any expectation value. We once again apply the doubling trick, taking into account the boundary changing operator that is contained in the junction and is not doubled. We obtain
\begin{equation}\label{eq:string-order-param}
   \mathcal{Z}_1(g) = \left \langle \,
    \tikz[baseline=.1ex,scale =0.8]{
    \draw[dashed] (-1.5,-1) rectangle (1.5,1);
    \draw[dotted] (0,-1) -- (0,1);
    \fill (1, 0) circle(1.5pt) node[below] {$\mu_\mathrm{c}$};
    \fill (-1, 0) circle(1.5pt) node[below] {$\mu_\mathrm{c}$};
    \fill (0,0) circle(1.5pt) node[below] {$\Psi_{(ga,a)}$};
    }
    \, \right \rangle_\mathrm{c}
    \propto \ell^{-2 h_\mu- h_{\Psi_{(ga,a)}}}
\end{equation}
where in the last step we used the expression for three point functions of primary fields in CFT and we have denoted by $h_{\Psi_{(ga,a)}}$ the scaling dimension of $\Psi_{(ga,a)}$. 
Therefore, the symmetry breaking boundary condition results in a stronger decay of the string order parameter
away from the boundary. As a consequence, for Lie groups this leads to larger fluctuations of the group generators in the subsystem, as one might expect. 

\subsection{Entanglement asymmetry}
We now turn to the calculation of the higher order normalized charged moments $\mathcal{Z}_n(\boldg)$ with a view to obtaining the entanglement asymmetry. In this section we restrict to the case where $G$ is  a discrete group and has cardinality $|G|$. Furthermore, we consider only a boundary condition which breaks $G$ fully, although the case where a subgroup is preserved can be studied using the same techniques. 
 
 We can express $\mathcal Z_n(\boldg)$ in terms of expectation values in the $n$-replicated theory of the group operator $U_{A,g_1} \otimes \dots \otimes U_{A,g_n}$ and the replica-change operator $R$, which we now introduce. Letting $T^{\otimes n}$ denote the $n$-replicated theory, 
 we take  $R$ be the line operator in $T^{\otimes n}$ that implements the $\ZZ_n$ cyclic permutation which shifts the replica labels by $1$. In other words, $R$ transforms a local operator, $\mathcal O = O_1 \otimes \dots \otimes O_n$ into $\mathcal O' = O_n \otimes O_1 \otimes \dots \otimes O_{n-1}$. This can be depicted graphically in the following way
\begin{equation}
    \tikz[baseline=.1ex,scale =0.7]
    {
    \draw[purple, thick] (0,0) -- (2,0) node[right] {$R$};
    \fill (1,-0.5) circle(1.5pt) node[left] {$\mathcal O$};
    }
    = 
    \tikz[baseline=.1ex,scale =0.7]
    {
    \draw[purple, thick] (0,0) -- (2,0) node[right] {$R$};
    \fill (1,0.5) circle(1.5pt) node[left] {$\mathcal O'$};
    }.
\end{equation}
An analogous action on line operators that cross $R$ also exists as does the natural graphical representation. We will call $R$ the \emph{replica change operator}. $R$ is the analogue of the branch cut of the Riemann surface when one replicates the space-time manifold instead of replicating the theory \cite{Calabrese:2004eu}. The line $R$ can end on a \emph{replica twist-field} $\mathcal T$ \cite{Calabrese:2004eu}, and when combined with a line operator that implements the element $\boldg$ can end on a  \emph{composite twist field} $\mathcal T^\boldg$, which is the one routinely used to compute symmetry resolved entanglement \cite{Goldstein:2017bua,Horvath:2020vzs}.

The normalized partition function, $\Zng$, can then be written as the ratio of two expectation values in the replicated theory.  Namely, it is   the ratio of the expectation value of the replicated line operator $U_{A,g_1} \otimes \dots \otimes U_{A,g_n}$ with that of the $R_A$, the replica line operator restricted to the subsystem.  In diagrams we express this as
\begin{equation}\label{eq:Zng-ratio-Zn}
    \Zng = \left. 
    \left \langle
    \tikz[baseline=.1ex,scale =0.8]{
        \draw[dashed] (0,-1) rectangle (2,1);
        \draw[very thick, orange] (0,-1) -- node[midway, left] {$a^{\otimes n}$} (0,1) ;
        
        \draw[thick, purple] (0,-0) -- node[midway, below] {$R$} (1,0);
        \draw[thick, blue] plot[smooth, tension=1] coordinates {(1,0) (0.8,0.1) (0,0.1) };
        \node[blue, above] at (0.5,0.1) {$\boldg$};
        \fill (1, 0) circle(1.5pt) node[right] {$\mathcal T^{\boldg }$};
    } \, 
    \right \rangle
    \, \middle/ \,
    \left \langle
    \tikz[baseline=.1ex,scale =0.8]{
        \draw[dashed] (0,-1) rectangle (2,1);
        \draw[very thick, orange] (0,-1) -- node[midway, left] {$a^{\otimes n}$} (0,1) ;
        
        \draw[thick, purple] (0,-0) -- node[midway, below] {$R$} (1,0);
        \fill (1, 0) circle(1.5pt) node[right] {$\mathcal T$};
    } \,
    \right \rangle
    \right.
    .
\end{equation}
To understand how we can evaluate this ratio we consider first,  the simplest case of $n=2$ and recall that in order to calculate $\Delta S_A$ we only require the neutral charged moments, i.e. $\mathcal{Z}_2(g,g^{-1})$.  In terms of these 
\begin{equation}\label{eq:asymmetry_n=2}
    \Delta S^{(2)}_{A} = \log |G|  - \log \lr{1+ \sum_{g \in G,g\neq e} \mathcal Z_2(g, g^{-1} ) },
\end{equation}
where we have used that $\mathcal{Z}_2(e,e)=1$.  To evaluate the elements of the sum, $\mathcal{Z}_2(g, g^{-1} ), ~g\neq e$ we make use of the fact that the line operator is topological in the bulk and also that we can move the lines in each replica independently. This means we can move the $g$ line in the first replica as in \eqref{eq:move-line-change-boundary} which leaves behind the boundary changing operator $\Psi_{(ga,a)}$ and creates a topological junction which can be moved at will. Since the points $\pm \infty$ in the temporal direction are identified, we can move the line $g$ along the whole first replica until it crosses the replica-change line $R$ from below. After crossing $R$, the line then switches to the second replica and the boundary condition on the first replica has changed to $ga$. The line $g$ then fuses with $g^{-1}$ in the second replica, producing the identity line but leaving behind its hidden boundary changing operator. The result is that the lines $g$ and $g^{-1}$ have disappeared,  the boundary condition on the first replica is $ga$, and that there are two boundary changing operators $\Psi_{(ga,a)}$, one for each replica. Pictorially, the result is
\begin{equation}\label{eq:Z2(etaeta)}
\begin{aligned}
        \mathcal Z_2(g, g^{-1}) &= 
        \left.  
        \left \langle
        \tikz[baseline=.1ex,scale =0.6]{
            \draw[dashed] (0,-2) rectangle (3,2);
            \draw[very thick, orange] (0,-2) -- node[midway, left] {$ga \otimes a$} (0,2) ;
            \draw[thick, purple] (0,-0) -- node[midway, below] {$R$} (1.8,0);
            \fill (1.8, 0) circle(2pt) node[right] {$\mathcal T$};
            \fill (0, 0) circle(2pt) node[above right] {$\Psi_{(ga,a)}^{\otimes 2}$};
        }  \, 
        \right \rangle 
        \, \middle/ \, 
        \left \langle
        \tikz[baseline=.1ex,scale =0.6]{
            \draw[dashed] (0,-2) rectangle (3,2);
            \draw[very thick, orange] (0,-2) -- node[midway, left] {$a^{\otimes 2}$} (0,2) ;
            
            \draw[thick, purple] (0,-0) -- node[midway, below] {$R$} (1.5,0);
            \fill (1.5, 0) circle(1.5pt) node[right] {$\mathcal T$};
        } \,
        \right \rangle
        \right. \\
        &= \frac{\langle \Psi_{(ga,a)}^{\otimes 2}(0) \mathcal T(\ell) \rangle }{\langle  \mathcal T(\ell) \rangle }.
\end{aligned}
\end{equation}
where in the second line we re-express the result in the more standard non-graphical manner. Then, using once again the doubling trick \cite{Cardy:1984bb}, this time for the replicated theory,  we have that $\langle \Psi_{(ga,a)}^{\otimes 2}(0) \mathcal T(\ell) \rangle/\langle  \mathcal T(\ell) \rangle$ is proportional to a ratio of chiral correlators in the replicated chiral theory $T_c^{\otimes n}$, i.e 
\begin{equation} \label{eq:doubling-trick-charged-moment}
\begin{aligned}
    \mathcal Z_2(g, g^{-1}) &=  
    \left .
    \left \langle \,
    \tikz[baseline=.1ex,scale =0.8]{
    	\draw[dashed] (-1.5,-1) rectangle (1.5,1);
    	\draw[dotted] (0,-1) -- (0,1);
    	\fill (1, 0) circle(1.5pt) node[above] {$\mathcal T_\mathrm{c}$};
    	\fill (-1, 0) circle(1.5pt) node[above] {$\bar{\mathcal {T}}_\mathrm{c}$};
    	\fill (0,0) circle(1.5pt) node[above] {$\Psi_{(ga,a)}^{\otimes 2}$};
    }
    \, \right \rangle_{\rm c}
    \, \middle/ \, 
     \left \langle \,
    \tikz[baseline=.1ex,scale =0.8]{
    	\draw[dashed] (-1.5,-1) rectangle (1.5,1);
    	\draw[dotted] (0,-1) -- (0,1);
    	\fill (1, 0) circle(1.5pt) node[above] {$\mathcal T_\mathrm{c}$};
    	\fill (-1, 0) circle(1.5pt) node[above] {$\bar{\mathcal{T}}_\mathrm{c}$};
    }
    \, \right \rangle_{\rm c}
     \right. 
     \\
     &=    \frac{\langle \bar{\mathcal{T}}_\mathrm{c} (-\ell) \Psi_{(ga,a)}^{\otimes 2}(0) \mathcal T_\mathrm{c} (\ell) \rangle_{\rm c} }{\langle  \bar{\mathcal{T}}_\mathrm{c} (-\ell) \mathcal{T}_\mathrm{c} (\ell) \rangle_{\rm c}}  .
\end{aligned}
\end{equation}
Since the structure of 3-point functions and 2-point functions of primary fields in the chiral theory are known we arrive at
\begin{equation}\label{eq:result-charged-moment}
	\mathcal Z_2 (g, g^{-1})  \propto  \ell^{ -2h_{\Psi_{(ga,a)}} }.
\end{equation}
Hence, we find that the  normalized charged moment decays algebraically in $\ell$ with a power that is two times the conformal weight of the boundary changing operator $\Psi_{(ga,a)}$. In contrast to the $n=1$ charged moment discussed in the previous subsection, the dimension of the defect operator $\mu$ does not appear, which results from the neutrality of the charged moment. Moreover, the dimension of the replica twist field operator is absent which originates from the normalization. 

Plugging Eq. \eqref{eq:result-charged-moment} into \cref{eq:asymmetry_n=2}, we can expand the $\log$ for large $\ell$. Retaining only the leading term in this expansion we get
\begin{equation}
    \Delta S^{(2)}_A = \log |G| - c_2 \ell^{-2h^\mathrm{min} } + o( \ell^{-2h^\mathrm{min} })
    \label{asyasy}
\end{equation}
with 
\begin{equation}
    h^\mathrm{min} = \min_{g \in G,g\neq e} h_{\Psi_{(ga,a)}}
\end{equation}
and $c_2$ is a proportionality constant that incorporates various non-universal terms and  the dependence on the short-distance regulator. 

The preceding analysis can be straightforwardly generalized to a  higher number of replicas. Consider the  normalized charged moment $\mathcal Z_n(\boldg)$ with $\boldg = (g_1, \dots, g_n)$, subject to the neutrality condition $\prod_{j=1}^n g_j = e$. We can repeat the topological manipulation we carried out above: The $g_1$ line is moved from the first replica to the second replica, leaving in its wake $\Psi_{(g_1a,a)}$. Then the boundary in the first replica becomes $g_1 a$ and the line in the second replica becomes $g_1 g_2$. The fused $g_1g_2$ line is then moved to the third replica, leaving behind this time $\Psi_{(g_1g_2a,g_1a)}$. Continuing in this fashion we fuse all the lines to the identity. In the end, we will get the following boundary condition of the replicated theory
\begin{equation}
    g_1 a \otimes g_1 g_2 a \dots \otimes \lr{ \prod_{j=1}^{n-1} g_j } \! a \otimes a
\end{equation}
and the insertion of following boundary changing operator
\begin{equation}
    \Phi = \Psi_{(g_1 a, a)} \otimes \Psi_{(g_1 g_2 a, g_1 a)} \otimes \dots  \otimes \Psi_{(a, (\prod_{j=1}^{n-1} g_j) a)}
\end{equation}
at the origin. 
The dimension $h_\Phi$ of $\Phi$ is just the sum of the dimensions of the boundary condition changing operators that comprise it.
After this, the doubling trick can be applied as in \eqref{eq:doubling-trick-charged-moment}, producing
\begin{equation}
    \mathcal Z_n(\boldg) \propto \ell^{- h_\Phi}.
\end{equation}
provided $\boldg\neq\boldsymbol{e}$ where $\boldsymbol{e}=(e,\dots,e)$ in which case $\mathcal{Z}_n(\boldsymbol{e})=1$. The leading behaviour in $\sum_{\boldg\neq \boldsymbol{e}} \mathcal Z_n(\boldg)$ for large $\ell$ is given by the values of $\boldg$ for which $h_\Phi$ is minimum. The minimum value of $h_\Phi$ is realized  only if $\boldg$ contains only two non-trivial group elements, and the other entries are identities, i.e. only if $\boldg$  is of the form $\boldg = ( \dots, g, \dots,  g^{-1}, \dots )$ where the dots are identities. Notice that this structure resembles  the one for the OPE of two twist fields where it enters the bulk operator with the minimal dimension \cite{Calabrese:2010he}. 

Consequently, the leading term in $\sum_{\boldg \neq \boldsymbol e} \mathcal Z_n(\boldg)$ goes like $\ell^{-2 h^\mathrm{min} }$ and in particular is independent of the replica index. 
From this we get that the leading asymptotic behavior of the R\'enyi entanglement asymmetry \emph{for every} $n$ is
\begin{equation}
    \Delta S^{(n)}_A = \log |G| - c_n \ell^{-2 h^\mathrm{min}} + o \lr{ \ell^{-2 h^\mathrm{min}} }.
\end{equation}
where $c_n$ is again a non-universal constant. Finally, to get the entanglement asymmetry, one should analytically continue the result in the variable $n$ and take the limit $n \to 1$. Since the power-law exponent is independent of $n$, we get
\begin{equation}
    \Delta S_A = \log |G| - c_1 \ell^{-2 h^\mathrm{min}} + o \lr{ \ell^{-2 h^\mathrm{min}} },
    \label{asyfin}
\end{equation}
where we have assumed that the analytic continuation of the amplitude, $c_1$, does not vanish. Thus, the entanglement asymmetry in a critical system with boundary symmetry breaking approaches a universal value, dependent only on the cardinality of the broken symmetry group, for large subsystem sizes. The corrections to this are algebraic with a universal exponent whose value is given by the smallest scaling dimension of the boundary changing operators which are not invariant under the symmetry. 

If the group $G$ is not fully broken by the boundary condition, leaving a residual symmetry subgroup $H$ of cardinality $|H|$, the asymmetry is given by Eq. \eqref{asyfin} with $|G|$ replaced by $|G|/|H|$ and now 
\begin{equation}
    h^\mathrm{min} = \min_{g \in G/H} h_{\Psi_{(ga,a)}}
\end{equation}
where $G/H$ denotes the elements which act non-trivially on $a$.

\section{Examples and numerical checks on the lattice}
\label{numerics}
In this section we examine our results on the entanglement asymmetry and the string order parameter $\mathcal{Z}_1(g)$ in two specific examples, the critical Ising model and the critical 3-state Potts model. We then check these predictions against numerical simulations on lattice systems which are described by these CFTs at low energy. 
All numerical simulations were conducted by constructing an MPS approximation of the ground state for a finite chain of total length $L$  using DMRG, implemented with the TeNPy library~\cite{Hauschild:2018kmz}.

\subsection{Critical Ising model}
We start with the Ising CFT which notoriously has an internal $\ZZ_2$ spin-flip symmetry. This is implemented by the Verlinde line associated to the $\epsilon$ primary field, that we denote by $\eta$. The conformal boundary conditions are denoted by $f, +, -$. Here, $f$ is the free boundary condition, where no constraint is set at the boundary spins, and $+, -$ are the fixed boundary conditions \cite{Cardy:1989ir}. 
When fused with the boundary as in Eq. \eqref{eq:move-line-change-boundary}, the $\eta$ line acts on the boundary conditions as $\eta f = f, \, \eta + = -, \, \eta - = +$. 
For this reason, when the $\eta$ line hits perpendicularly the boundary, it forms a topological junction with the $f$ boundary condition. 
In contrast, the junction with a fixed boundary condition is necessarily non-topological. 
Moving the $\eta$ line when attached to the $+$ boundary as in \eqref{eq:move-line-change-boundary} produces a boundary changing operator $\Psi_{(+,-)}$, which has conformal dimension $h_{\Psi_{(+,-)}}=1/2$ \cite{Cardy:1986ie}. 
In the bulk, the $\eta$ line can end on the primaries $\mu, \psi, \bar \psi$  \cite{Petkova:2000ip, Shao:2023gho}. 

The Ising CFT is realized as the low energy effective field theory of the critical transverse field Ising chain. The Hamiltonian of the semi-infinite chain with free boundary condition at its left edge is given by
\begin{equation}\label{eq:H_Ising_free}
	H_f =  - \sum_{j=1}^\infty ( X_j X_{j+1} + Z_j ),
\end{equation}
where $X_j,Z_j$ are the Pauli matrices for the spins at site $j$ and the free boundary is simply implemented by truncating the chain at site $j=1$. We denote the ground state of this system by $\ket{\mathrm{gs},f}$. To implement the fixed boundary conditions, however we should include an extra site at $j=0$ on which an additional longitudinal field acts. The Hamiltonian is 
\begin{equation}\label{eq:H_Ising_fixed}
    H_{\pm} =  - \sum_{j=0}^\infty ( X_j X_{j+1} + Z_j ) - h X_0.
\end{equation}
where the $\pm$ boundary conditions arise in the limit $h\to\pm\infty$. In this limit, the $j=0$ spin  becomes completely magnetized, and any finite energy state is of the form $\ket \psi = \ket \pm \otimes \ket{\text{bulk}}$, where $\ket{\pm}$ denote the eigenstates of $X_0$ and $\ket{\rm bulk}$ denotes the state of the system for $j\geq 1$. It is important to note that the site at $j=0$ is an auxiliary spin which is included so as to implement the fixed boundary condition. At low energy it is the properties of the $\ket{\rm bulk}$ part of the states which will be described by the Ising CFT. 
We denote the ground state of this Hamiltonian, including the auxiliary spin, by $\ket{\mathrm{gs},\pm}$. 

\begin{figure}
    \centering
     \includegraphics[width=0.5\textwidth]{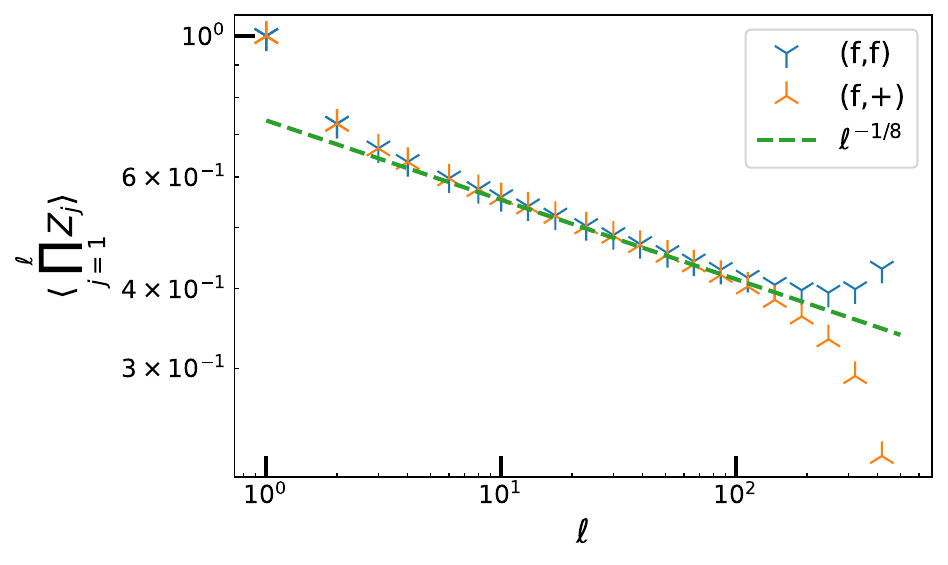}
    \includegraphics[width=0.48\textwidth]{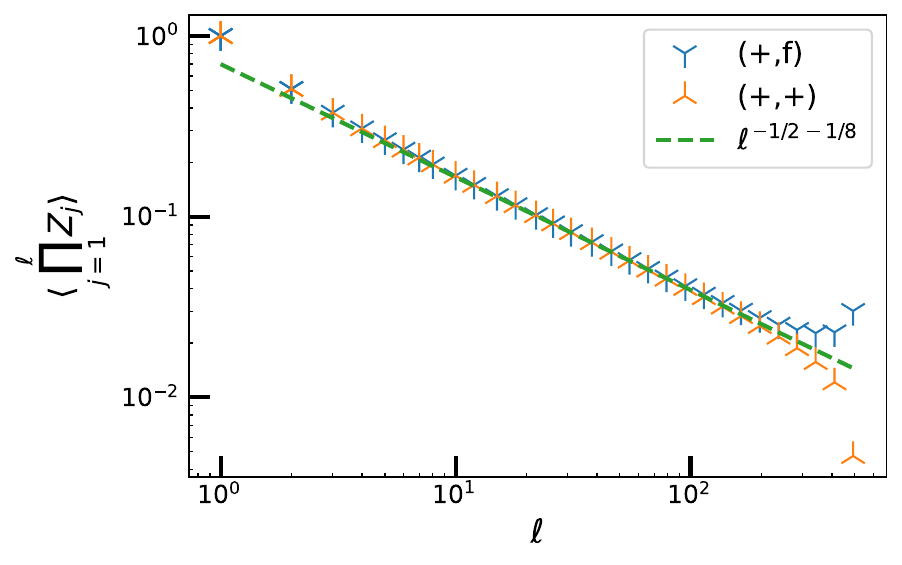}
    \caption{
    Expectation value of the string order parameter $\eta_{A}= \prod_{j \in A} Z_j$ in the Ising chain with different BCs.  On the left we set the BC in $0$ to be $f$, and consider the combinations $(f,+)$ and $(f,f)$. We observe excellent agreement with the predicted decay, which scales as  $\ell^{-1/8}$ \eqref{eq:exp_val_flip_f}. On the right, we set the BC in $0$ to be $+$, and we consider the combinations $(+,f)$ and $(+,+)$. We observe good agreement with the prediction of a decay proportional to $\ell^{-5/8}$, cf. \eqref{eq:exp_val_flip_+}. 
    The spin chain has $L=500$ sites, and the $+$ BC is implemented with a boundary field $h = 500$. 
    }
    \label{fig:flip_exp_val}
\end{figure}

\paragraph{The string order parameter. } The string order parameter is given by the expectation value of the lattice spin flip operator on the subsystem $A$ which is $\eta_A = \prod_{j \in A} Z_j$. Under the renormalization group, the bulk endpoint becomes the disorder operator $\mu$ with conformal dimension $(1/16, 1/16)$. The primaries $\psi$ and $\bar \psi$ can be ruled out as  they are $\ZZ_2$-odd. 
If the lattice boundary condition is free, the operator $\prod_{j=1}^\ell Z_j$ flows to the $\eta$ line, with the topological junction with the free boundary. 
The analytical prediction in Eq. \eqref{eq:expectation-value-line-invariant-boundary}  gives 
\begin{equation} \label{eq:exp_val_flip_f}
	\bra{ \mathrm{gs}, f} \eta_A \ket{\mathrm{gs}, f} \propto \ell^{- 1/8}. 
\end{equation}
To check this prediction we perform numerical simulations using tensor networks on the Hamiltonian~\eqref{eq:H_Ising_free}. To implement this we use a finite but large chain of $L=500$ sites and impose either free or $+$ boundary conditions at the site $j=L$. The results are shown in the left panel of Eq.~\cref{fig:flip_exp_val} where the symbols indicate numerical data and the dashed line depicts the prediction of~\eqref{eq:exp_val_flip_f}. We see excellent agreement up to subsystem sizes approaching half the chain length when finite size effects become important.

We now examine the case of fixed boundary condition, focusing on the $+$ case. In this instance, the operator $\prod_{j=1}^\ell Z_j$ flows to the $\eta$ line connected with the non-topological junction to the  $+$ boundary. 
We stress that here we did not include the site $0$ in the spin flip operator. Indeed  $\prod_{j=0}^\ell Z_j$ flips the site $0$ and therefore is a map from the Hilbert space with $+$ bc to the Hilbert space with $-$ bc. Therefore, $\prod_{j=0}^\ell Z_j$  flows to the $\eta$ line attached to the boundaries $+$ and $-$ with a topological junction. This can be summarized graphically as
\begin{align}
	\prod_{j=1}^\ell Z_j \leadsto  \tikz[baseline=.1ex,scale =0.7]{
	    \draw[very thick, orange] (0,-1) -- node[midway, left] {$+$} (0,1) ;
    \draw[thick, blue] (0,0) -- node[midway, above] {$\eta$} (2,0);
    \fill (0,0) circle (2pt);
    \fill (2, 0) circle(1.5pt) node[right] {$\mu$};
    }
    \qquad \quad
    \prod_{j=0}^\ell Z_j 
    \leadsto	\tikz[baseline=.1ex,scale =0.7]{
    \draw[very thick, orange] (0,-1) -- node[midway, left] {$+$} (0,0) ;
    \draw[very thick, orange] (0,0) -- node[midway, left] {$-$} (0,1) ;
    \draw[thick, blue] (0,0) -- node[midway, above] {$\eta$} (2,0);
    \fill (2, 0) circle(1.5pt) node[right] {$\mu$};
    }.
\end{align}
As a result one can immediately determine that
\begin{equation}
    \bra{\mathrm{gs}, +} \prod_{j=0}^\ell Z_j \ket{\mathrm{gs}, +} = 0.
\end{equation}
On the other hand, a sensible choice is to restrict the flip operator to the bulk sites, excluding the frozen spin at $j=0$. In that case, our prediction~\eqref{eq:string-order-param} reads,
\begin{equation} \label{eq:exp_val_flip_+}
	\bra{\mathrm{gs}, +} \prod_{j=1}^\ell Z_j \ket{\mathrm{gs}, +} \propto \ell^{- 1/8 - 1/2}=\ell^{- 5/8}. 
\end{equation}
We check this numerically, as before and plot the result in the right panel of ~\cref{fig:flip_exp_val}.  Once again we see excellent agreement between numerics and the CFT prediction. 

\begin{figure}
    \centering
    \includegraphics[width=0.49\textwidth]{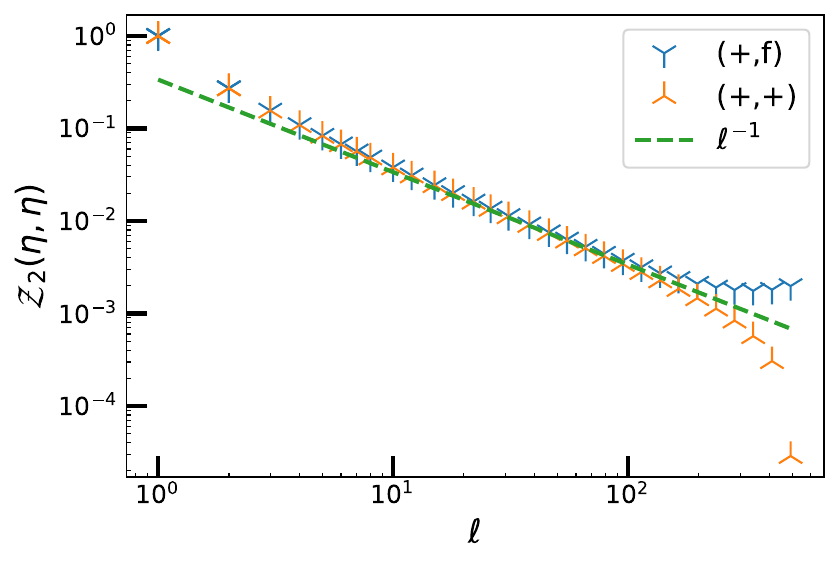}
    \includegraphics[width=0.49\linewidth]{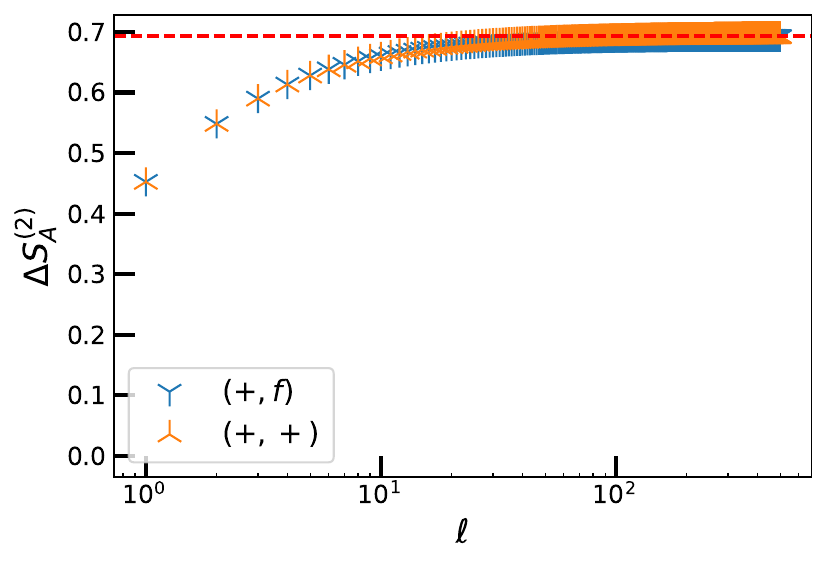}
    \caption{Normalized charged moment $\mathcal Z_2(\eta, \eta)$ (left) and second R\'enyi entanglement asymmetry $\Delta S^{(2)}_A$ (right) for the critical Ising model with $+$ boundary condition imposed at $j=1$. 
    The parameters are the same as in \cref{fig:flip_exp_val}. 
    On the left, we display $\mathcal Z_2(\eta, \eta)$ and  the dashed line is  the prediction ~\eqref{eq:charged_moment_pred_Ising}. 
    On the right, we plot the entanglement asymmetry which approaches $\log2$, independently of the boundary condition imposed at the right boundary.}      \label{fig:Z2(etaeta)_numerics}
\end{figure}

\paragraph{Entanglement asymmetry.}
We now determine the $\ZZ_2$ entanglement asymmetry for the Ising model with $+$ boundary condition.  
We specifically investigate the $n=2$ case, which is in fact the most experimentally relevant instance~\cite{Joshi:2024sup}. 
In this case, the asymmetry is given exactly by 
\begin{equation}
    \Delta S^{(2)}_A = \log 2 - \log \lr{ 1 + \mathcal Z_2(\eta, \eta)},
\end{equation}
where the CFT predicts that charged moment behaves as 
\begin{equation}\label{eq:charged_moment_pred_Ising}
    \mathcal Z_2 (\eta, \eta) \propto \ell^{- 1} 
\end{equation}
and we used that $h_{\Psi_{(+,-)}}=1/2$ in Eq. \eqref{asyasy}.  
In the left panel of~\cref{fig:Z2(etaeta)_numerics}, we compare our prediction for the charged moment with numerical data, which confirms the accuracy of the theoretical result: 
the correct power law scaling is reproduced before finite size effects kick in. 
In the right panel we plot the entanglement asymmetry which exhibits the anticipated behaviour: $\Delta S_A$ approaches $\log 2$ with corrections going as $1/\ell$ for large subsystem size. 

\subsection{Critical 3-state Potts model}
Having examined the simplest case of a CFT with a discrete symmetry in the previous section, we now move on to a more intricate example and study the $n=2$ entanglement asymmetry for the $\ZZ_3$ symmetry of the critical three-state Potts model, in the case of fixed boundary conditions.

The three state Potts model is a generalization of the Ising model with an enlarged global symmetry \cite{DiFrancesco:1997nk}. A possible lattice realization is made in terms of a spin chain chain with a three-dimensional local Hilbert space $\mathbb C^3$. 
The critical Hamiltonian on a semi infinite chain with a free boundary condition on the left edge is given by 
\begin{equation}\label{eq:potts-hamiltonian-free}
    H_f = - \sum_{j=1}^\infty ( X_j X_{j+1}^\dagger + X_{j+1} X_j^\dagger + Z_j + Z_j^\dagger )
\end{equation} where we have introduced  the clock operator $X$ and the shift operator $Z$, defined by
\begin{align}
    X &= 
    \begin{pmatrix}
        0 & 1 & 0 \\
        0 & 0 & 1 \\
        1 & 0 & 0
    \end{pmatrix}
    &
    Z &= \begin{pmatrix}
        1 & 0 & 0 \\
        0 & e^{i 2\pi/3} & 0 \\
        0 & 0 & e^{-i 2\pi/3}
    \end{pmatrix}.
\end{align}
This lattice model flows to the three-state Potts CFT which admits eight conformal boundary conditions \cite{Affleck:1998nq, Fuchs:1998qn}. The simplest ones to implement on the lattice are the free and the three fixed ones. The free is implemented by simply truncating the chain as is done in \eqref{eq:potts-hamiltonian-free}. The fixed boundary conditions can be implemented in the lattice using a similar procedure to the Ising model.  We add an  auxiliary spin to the chain and fix it to be in one of the states $\ket{A}=(1,1,1)^T, \ket{B}=(1,e^{i2\pi/3},e^{-i2\pi/3})^T$ or $\ket{C}=(1,e^{-i2\pi/3},e^{i2\pi/3})^T$ with an energetic constraint which is imposed by an external field. For instance, the fixed $A$-type boundary condition can be obtained with the Hamiltonian
\begin{equation}
    H_A = - \sum_{j=0}^\infty ( X_j X_{j+1}^\dagger + X_{j+1} X_j^\dagger + Z_j + Z_j^\dagger ) - h (X_0 + X_0^\dagger)
\end{equation}
in the limit $h \to +\infty$. As before, we should consider only the sites $j\geq 1$ to be the region described by the CFT. 
The three-state Potts model has an internal $S_3$ symmetry and we focus on the breaking of its $\ZZ_3$ Abelian subgroup by the inclusion of the $A$-type boundary condition.  This $\ZZ_3$ symmetry is generated by
\begin{equation}
    \eta_A = \prod_{j=1}^\ell Z_j
\end{equation}
which acts by permuting cyclically $A, B$ and $C$. In the bulk, this operator flows to the topological line $\eta$ (notation from \cite{Chang:2018iay}). The junction with the $A$-type boundary ``contains'' a boundary-changing operator $\Psi_{(B,A)}$, as one can see from
\begin{equation}
    \tikz[baseline=.1ex,scale =0.7]{
    \draw[very thick, orange] (0,-1) -- node[midway, left] {$A$} (0,1) ;
    \draw[thick, blue] (0,-0.5) -- node[midway, above] {$\eta$} (2,-0.5);
    \fill (0,-0.5) circle (2pt);
    }
    =
     \tikz[baseline=.1ex,scale =0.7]{
     \draw[very thick, orange] (0,-1) -- node[midway, left] {$A$} (0,1) ;
    \draw[thick, blue] plot[smooth, tension=1] coordinates {(0,-0.5) (0.2,-0.3) (0.45,0.5) (2,-0.5)};
    \fill (0,-0.5) circle (2pt);
    }
    =
    \tikz[baseline=.1ex,scale =0.7]{
    \draw[very thick, orange] (0,-1) -- node[midway, left] {$A$} (0,-0.5) -- node[midway, left] {$B$} (0,0.5) -- node[midway, left] {$A$} (0,1) ;
    \draw[thick, blue] (0,0.5) -- node[midway, above right] {$\eta$} (2,-0.5) ;
    \fill (0,-0.5) circle (2pt) node[right] {$\Psi_{(B, A)}$};
    },
\end{equation}
and $\Psi_{(B,A)}$ has conformal dimension $2/3$ \cite{Cardy:1989ir}.

\begin{figure}
    \centering
    \includegraphics[width=0.49\linewidth]{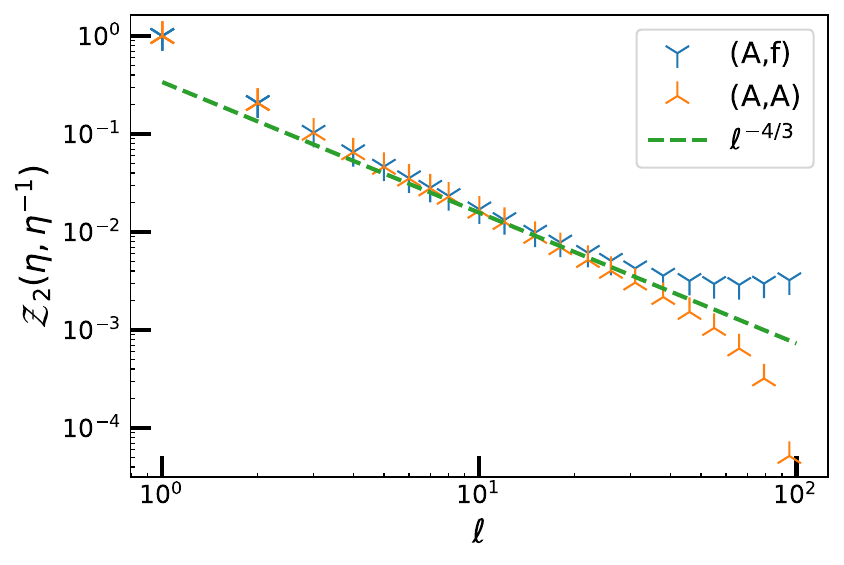}
    \includegraphics[width=0.49\linewidth]{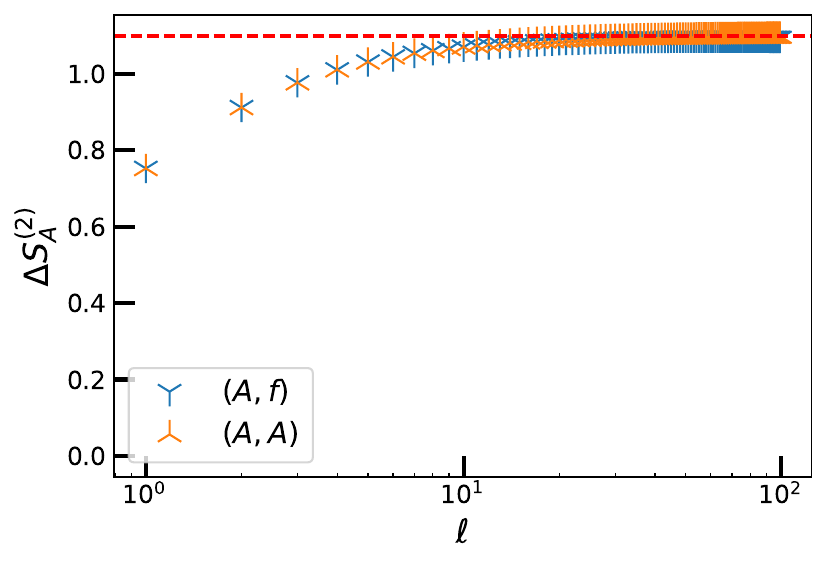}
    
    \caption{Normalized charged moment $\mathcal Z_2(\eta, \eta^{-1})$ (left) and second R\'enyi entanglement asymmetry $\Delta S^{(2)}_A$ (right) for the critical 3-state Potts model with the $A$-type boundary condition imposed at $j=1$. 
    The data corresponds to chain of total size $L=100$.  
    On the left, we display $\mathcal Z_2(\eta, \eta^{-1})$ and  the dashed line is  the prediction ~\eqref{asyz2}. 
    On the right, we plot the entanglement asymmetry which approaches $\log3$, independently of the boundary condition imposed at the right boundary.}
    \label{fig:potts-charged-moment}
\end{figure}

From the definition \eqref{eq:asymmetry-sum-group}, the second R\'enyi asymmetry for the $\ZZ_3$ group is
\begin{equation}
    \Delta S^{(2)}_A = \log 3 - \log \lr{ 1 + \mathcal Z_2(\eta, \eta^{-1}) + \mathcal Z_2(\eta^{-1}, \eta) },
   \label{asyz3} 
\end{equation}
and from the cyclicity of the trace in \eqref{eq:def-normalized-charged-moment}, it follows that $\mathcal Z_2(\eta, \eta^{-1}) = \mathcal Z_2(\eta^{-1}, \eta)$. In the large $\ell$ limit, the lattice model is described by the CFT, where, according to \eqref{eq:result-charged-moment}, we have 
\begin{equation}
    \mathcal Z_2(\eta, \eta^{-1}) \propto \ell^{-4/3}.
    \label{asyz2}
\end{equation} 
This result is numerically validated in \cref{fig:potts-charged-moment}. In the left panel, we display data obtained for a chain of size  $L = 100$. The numerical results align closely with the theoretical prediction given by Eq. \eqref{asyz2}, demonstrating both the accuracy of our analytical approach and the robustness of the numerical methods employed. 

Finally, using Eq. \eqref{asyz3} we have that the asymmetry approaches $\log 3$ with a power law correction given by
\begin{equation}\label{eq:asymm_potss_pred}
    \Delta S_A^{(2)} = \log 3 - c_2 \ell^{-4/3} + o\lr{ \ell^{-4/3} }.
\end{equation}
On the right of  \cref{fig:potts-charged-moment} we plot the entanglement asymmetry obtained numerically and see that it behaves as predicted by~\eqref{eq:asymm_potss_pred}.

\subsection{Interval starting at the right boundary}
In the preceding sections, we examined symmetry breaking in a subsystem that starts at the symmetry-breaking boundary condition. However, it is also insightful to explore the case where the subsystem does not include this boundary. 
Unlike entanglement entropy, entanglement asymmetry is not symmetric with respect to exchanging the subsystem with its complement, i.e. $\Delta S_A\neq \Delta S_{\bar{A}}$ where $\bar{A}$ is the set $x\notin A$. 
This is natural since there is no reason to expect, e.g., the level of symmetry breaking in a small subsystem, $A$ be related to the that of the much bigger $\bar A$. 
To explore this further we consider numerical simulations in both the 
critical Ising and three-state Potts chains of the normalized charged moments and entanglement asymmetry for $\bar{A}$.
For this geometry we have no CFT prediction.

\begin{figure}
    \centering
    \includegraphics[width=0.49\linewidth]{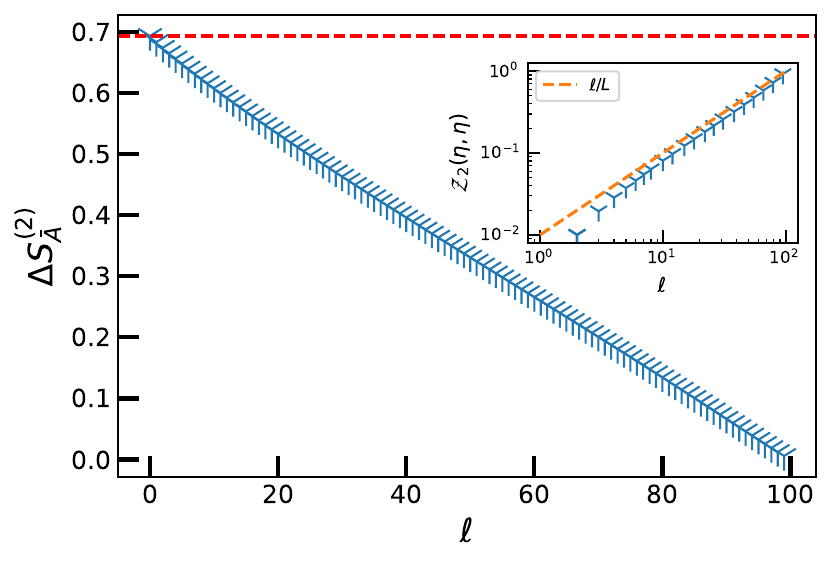}
    \includegraphics[width=0.49\linewidth]{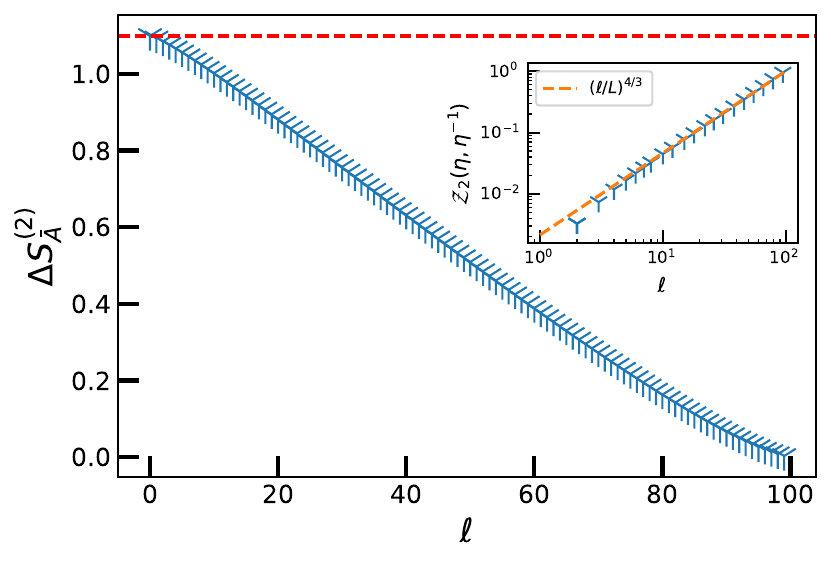}
    
    \caption{The entanglement asymmetry $\Delta S_{\bar A}$ for the subsystem $\bar{A}=(\ell,L]$. The insets show the normalized charged moments, $\mathcal{Z}_2$. 
    Left: The ground state of the Ising model with $+$ boundary condition at $j=1$ and free boundary condition at $j=L$. The parameters are $L=100$, $h=500$. The red dashed line indicates the maximum value of $\log 2$.
    In the inset the dashed line is the power law behavior $\ell/L$ without any fitting parameter. 
    Right:  The ground state of the 3-state Potts model with $A$ boundary condition at $j=1$ and free boundary condition at $j=L$. The parameters are $L=100$, $h=500$. The red dashed line indicates the maximum value of $\log 3$.
    In the inset the dashed line is the power law behavior $(\ell/L)^{4/3}$ without any fitting parameter. }
    \label{fig:right_boundary}
\end{figure}

Let us start with a finite critical Ising chain with $+$ and free boundary conditions at either end.  We first examine the normalized charged moment ${\cal Z}_2(\eta,\eta)$ which is reported in the inset of the left panel of Fig. \ref{fig:right_boundary}.
It turns out that this is well described by the curve $\ell/L$ without any adjustable parameter.
Then we move to the entanglement asymmetry $\Delta S_{\bar A}$ reported in the main panel. We observe an almost linear behavior; however, upon closer inspection, a slight bending becomes apparent.
Notice that for $\ell=0$, we recover $\log2$, as it should, since there we probe the symmetry breaking of the full system.
The behavior of both the charged moment and the entanglement asymmetry
is markedly different from the behavior seen above in~\cref{fig:Z2(etaeta)_numerics}.

We now move to the  critical 3-state Potts chain with the $A$-type and free boundary conditions at either end. 
We first examine the normalized charged moment ${\cal Z}_2(\eta,\eta^{-1})$ which is reported in the inset of the right panel of Fig. \ref{fig:right_boundary}.
It turns out that this is well described by the curve $(\ell/L)^{4/3}$ without any adjustable parameter.
The entanglement asymmetry $\Delta S_{\bar A}$ is reported in the main panel. 
Again we have  an almost linear behavior, but this time the bending is much more pronounced compared to the Ising model on the left.
For $\ell=0$, we recover $\log3$, as expected.

Overall, these numerical results indicate that the normalized charged moments exhibit algebraic scaling, with an exponent that may be connected to the dimension of the boundary condition changing operator. However, we currently lack a concrete CFT prediction for this specific geometry. Nonetheless, the simplicity of the observed behavior suggests that deriving a corresponding CFT prediction for this scenario should be feasible, but it is beyond the scope of this paper.

\section{Conclusions}
\label{concl}
In this paper, we have investigated explicit boundary symmetry breaking in $(1+1)-$ dimensional CFTs through the lens of the entanglement asymmetry. 
Using  the framework of boundary conformal field theory, we calculated both the entanglement asymmetry, $\Delta S_A$ and the string order parameter, a quantity closely related to the full counting statistics, in ground states of $G$-symmetric CFT.
These computations specifically addressed scenarios where the symmetry is broken by boundary conditions that are not $G$-invariant. 
We obtained a universal expression for the normalized charged moments, a central component to the calculation of $\Delta S_A$. This expression reveals  that charged moments decay algebraically with an exponent given by the dimension of the boundary condition changing operators of the theory. From this result, we concluded that the asymmetry approaches a universal constant, $\log|G|$,  with power-law corrections. The correction exponent is twice the smallest scaling dimension of the boundary condition changing operators.   
We then checked these predictions against numerical data for two lattice models whose low energy description is given by the Ising and 3-state Potts CFTs.
The numerical results demonstrated excellent agreement with our theoretical predictions.

Our work sheds further light  on the topic of symmetry breaking in many body quantum systems using methods other than the traditional order parameter approach. As a first case study, we have focused on the ground state properties of $(1+1)$-
dimensional CFTs with invertible symmetries, however, natural follow ups arise by relaxing any of these properties. In particular, the notion of non-invertible symmetries and their interplay with entanglement entropy has attracted an increasing amount of attention recently~\cite{Benedetti:2024dku,Heymann:2024vvf,Choi:2024wfm,Das:2024qdx}. Extending the framework presented here to incorporate boundary non-invertible symmetry breaking is therefore an intriguing avenue to explore. Alternatively, one can examine the fate of boundary symmetry breaking  after a quench to the symmetry preserving boundary condition. We will report on these aspects elsewhere~\cite{Fossati:2024quench}.  
Finally, it would be highly instructive to explore boundary symmetry breaking in non-critical (massive) quantum field theories. For instance, adapting the form factor approach of Ref. \cite{CapizziMazzoniIsing:2023} to the boundary case could yield valuable insights.

\section*{Acknowledgements} 
We thank Yuya Kusuki, Sara Murciano, Hirosi Ooguri, and Sridip Pal for sharing the results of their related work with us. Their findings, obtained through a different approach, are equivalent to ours \cite{Kusuki:2024abc}.
We thank Eran Sela, Eytan Grosfeld, Filiberto Ares, and especially Andrei Rotaru for discussions and collaboration on related topics. We thank Marina Huerta for bringing references~\cite{Casini:2019kex,Casini:2020rgj,Magan:2021myk,Benedetti:2024dku} to our attention. 
This research was supported by the European Research Council under Consolidator Grant number 771536 ``NEMO''.
PC and CR also acknowledge the  European Union - NextGenerationEU, in the framework of the PRIN Project HIGHEST code 2022SJCKAH.

\bibliographystyle{ytphys}
\bibliography{bibliography}

\end{document}